\documentclass[a4paper, 10pt, twocolumn]{article}%% for arxiv

\usepackage[utf8]{inputenc} % allow utf-8 input
\usepackage[T1]{fontenc}    % use 8-bit T1 fonts
\usepackage{authblk}
\usepackage{hyperref}       % hyperlinks
\usepackage{url}            % simple URL typesetting
\usepackage{booktabs}       % professional-quality tables
\usepackage{amsfonts}       % blackboard math symbols
\usepackage{nicefrac}       % compact symbols for 1/2, etc.
\usepackage{microtype}      % microtypography
\usepackage{lipsum}		% Can be removed after putting your text content
\usepackage{graphicx}
\usepackage{natbib}
\usepackage{doi}
\usepackage{tikz, tikz-qtree}
\usepackage{qcircuit}
\usepackage{braket}
\usepackage{amsmath,blkarray,booktabs, bigstrut}
\usepackage{algorithm}
\usepackage{algpseudocode}

\usepackage[hmargin=0.5in,vmargin=0.9in]{geometry}

\title{Quantum Radio Astronomy: \\ Quantum Linear Solvers for Redundant Baseline Calibration\thanks{preprint submitted to Astronomy and Computing}}

\author[1,2]{Nicolas Renaud\thanks{Corresponding author: \texttt{n.renaud@esciencecenter.nl},}}
\author[1]{Pablo Rodr\'iguez-S\'anchez}
\author[1]{Johan Hidding}
\author[3]{P. Chris Broekema}

\affil[1]{Netherlands eScience Center, Amsterdam, the Netherlands}
\affil[2]{Quantum Application Lab, Amsterdam, the Netherlands}
\affil[3]{Netherlands Institute for radio astronomy (ASTRON), Dwingeloo, the Netherlands}

\date{}

\begin{document}

\maketitle 

%\begin{frontmatter}

%%\title{Integrating Quantum Computers in Redundant Calibration Pipelines for Radio Astronomy}
% \title{Variational Quantum Linear Solver for Redundancy Calibration in Radio Astronomy on a 16 qubits IBM quantum computer.}
%\title{Quantum Radio Astronomy: \\ Quantum Linear Solvers for Redundant Baseline Calibration}
%\date{September 9, 1985}	% Here you can change the date presented in the paper title
%\date{} 					% Or removing it

%\author[1,2]{Nicolas Renaud\corref{cor1}\fnref{fn1}}
%\ead{n.renaud@esciencecenter.nl}
%\author[1]{Pablo Rodr\'iguez-S\'anchez}
%\ead{p.rodriguez-sanchez@esciencecenter.nl}
%\author[1]{Johan Hidding}
%\ead{j.hidding@esciencecenter.nl}
%\author[3]{P. Chris Broekema}
%\ead{broekema@astron.nl}

%% \affiliation[1]{organization={Netherlands eScience Center},
%% addressline={Matrix Building III},
%% city={Amsterdam},
%% country={The Netherlands}}
%% \affiliation[2]{organization={Quantum Application Lab},
%% addressline={Science Park, Startup Village},
%% city={Amsterdam},
%% country={The Netherlands}}
%% \affiliation[3]{organization={Netherlands Institute for radio astronomy (ASTRON)},
%% addressline={Oude Hoogeveensedijk 4},
%% city={Dwingeloo},
%% country={the Netherlands}}

%% \fntext[fn1]{Corresponding author}

\begin{abstract}
  The computational requirements of future large scale radio telescopes are expected to scale well beyond the capabilities of conventional digital resources. Current and planned telescopes are generally limited in their scientific potential by their ability to efficiently process the vast volumes of generated data. To mitigate this problem, we investigate the viability of emerging quantum computers for radio astronomy applications. In this a paper we demonstrate the potential use of variational quantum linear solvers in Noisy Intermediate Scale Quantum (NISQ) computers and combinatorial solvers in quantum annealers for a radio astronomy calibration pipeline. While we demonstrate that these approaches can lead to satisfying results when integrated in calibration pipelines, we show that current restrictions of quantum hardware limit their applicability and performance.  
\end{abstract}

% not needed due to use of frontmatter (which is needed to include keywords)
%\maketitle

% keywords can be removed
%% \begin{keyword}
%% quantum computing \sep 
%% quantum annealing \sep
%% instrumentation: interferometers \sep 
%% techniques: interferometric \sep 
%% radio continuum: general \sep
%% \end{keyword}

%\end{frontmatter}

\section{Introduction}

\begin{figure}
    \centering
    \includegraphics[scale=0.35]{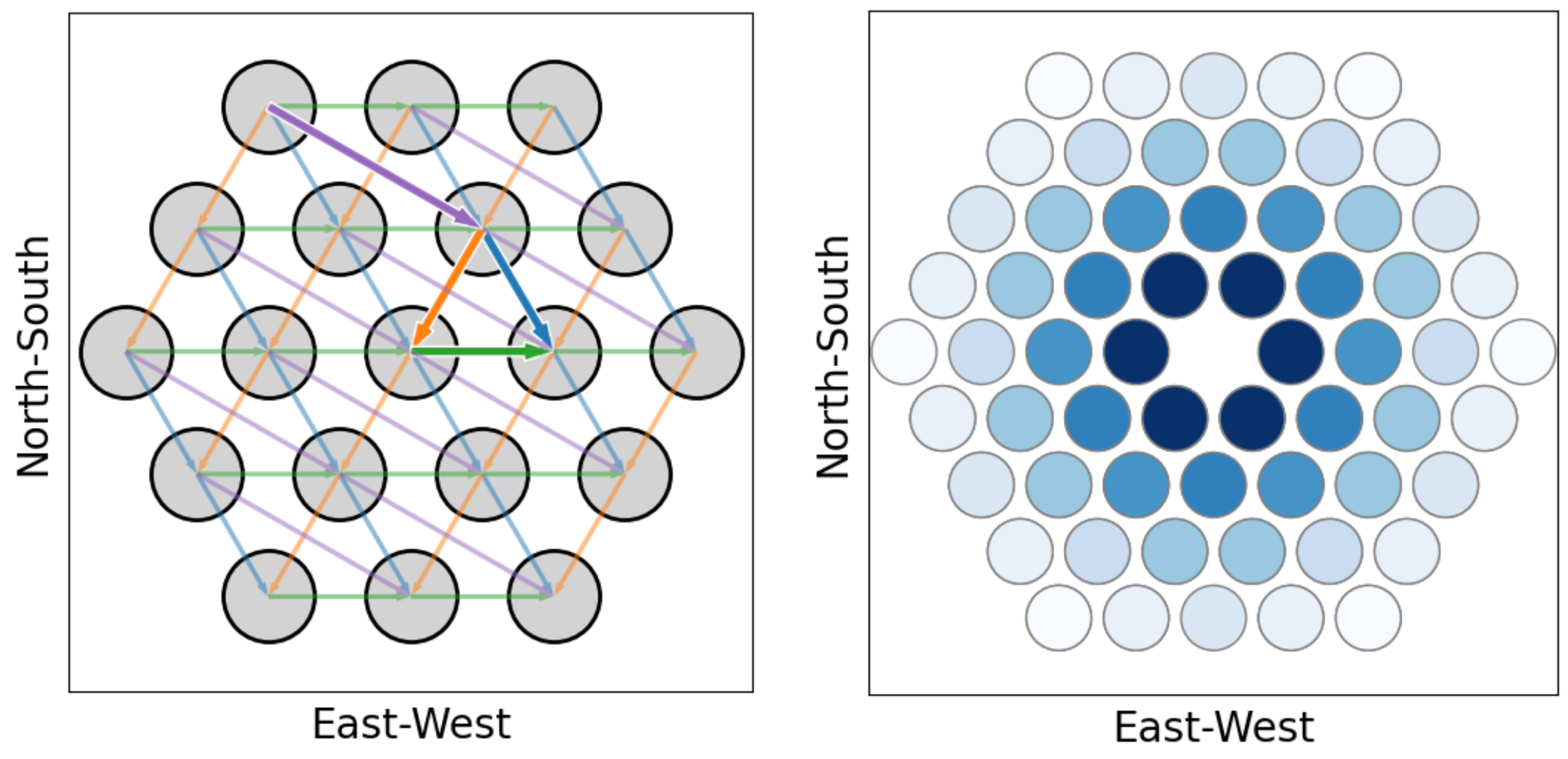}
    \caption{Left: Positions of the 17 antennas of a small hexagonal array. The bold arrows represent some of the unique baseline, while he thin ones mark the corresponding redundant baseline families. Right: Distribution of the unique baseline displacement color coded by the number of corresponding redundant baselines.}
    \label{fig:hera_guada}
\end{figure}

Calibration of large scale radio telescopes is computationally expensive and new methods need to be developed to meet the computational demand as well as to lower down the energy requirement.
One class of calibration exploits redundancy introduced by regular arrays to self-calibrate the array in a statistically efficient manner~\citep{wijnholds:2012}.
We explore a quantum accelerated version of this form of calibration due to its fairly simple structure, yet real-world applicability.
Furthermore, redundancy calibration relies heavily on solving sets of linear equations, for which efficient quantum algorithms are known to exist.
One radio telescope that utilises a very regular array structure is HERA, the Hydrogen Epoch of Reionization Array. A representation of the antennas position of the HERA telescope is show in Fig. \ref{fig:hera_guada}. The telescope is composed of a large number of antennas in a regular hexagonal pattern. This configuration leads to high degree of redundancy between baselines and is therefore well suited for redundancy calibration \citep{DeBoer_2017}.

The landscape of quantum computing has drastically changed over the last decade. While restricted to pen and paper studies and few qubits experimental implementation up until the early 2000s, quantum computers with several hundreds of qubits are now readily accessible through the cloud for researchers to experiment with. Several providers of quantum computers have emerged, each with their own specific technology \citep{resch2019quantum, qccloud, qc_review}. It is for example now possible to access quantum computers with 400+ superconducting qubits through the IBM cloud service\footnote{\url{https://quantum-computing.ibm.com}}. The accessibility of such large quantum computers has given rise to an rapidly growing number of applications, some of which claim to  already be outperforming classical computers \citep{supremacy, utility, Pokharel_2023, Denchev_2016}. While these claims have drawn significant criticism \citep{troyer_dwave_nospeedup, ChanUtility}, it is crucial to assess how quantum computers can be integrated in scientific computing pipelines and explore the benefit they might provide \citep{quantum_utility_definition}. In this article we explore how to leverage two distinct approaches to quantum computing, namely gate-based quantum computers such as the ones provided by IBM, and quantum annealers such as those developed by D-Wave, for calibration: a computationally expensive part of standard radio astronomy processing pipelines that require solving sets of linear equations. 

Several methods have been developed to solve linear systems using gate-based quantum computers. 
One of the most promising algorithm, the Harrow-Hassidim-Lloyd (HHL) method \citep{Harrow_2009, hhl_stepbystep}, promises a significant speed up compared to classical approaches. 
However, this method suffers from several caveats, both with respect hardware limitation, i.e. the unavailability of Quantum RAM, as well as with boundary conditions on the input data, i.e. the input matrix $A$ needs to be both sparse and \textit{Robustly invertible}. HHL is therefore currently impractical \citep{read_fine}. 
As an alternative we explore an variational approach that is better suited for current hardware, namely the Variational Quantum Linear Solver (VQLS) \citep{vqls}. 
New variations of the original VQLS method have been proposed to alleviate some of its limitations \citep{vqls_2} and it has been exploited to solve finite-element problems \citep{vqls_fe, vqls_flow}.

Quantum annealers (QAs) offer an interesting alternative to gate based quantum computers that has been extensively explored for real world applications \citep{QA_industry}. QA are a specialized type of hardware very suitable to solve minimization problems. As such they have been used to solve scheduling \citep{qa_transport} and power grid management \citep{qa_powergrid} problems but also more fundamental studies in geo-acoustic \citep{qa_seismic} and structural biology \citep{qa_peptide}. These studies leveraged D-Wave quantum annealers containing more than 2000 qubits and accessible through the cloud. While more suited to solve binary problems, QA are also capable of performing floating point calculation \citep{FloatQuantAn} and solving linear systems \citep{ViableQUBOLS}.

While we aim to minimize the quantum and classical computational cost of our simulations in this work, our goal is not to demonstrate quantum advantage or speed up. Instead we are attempting to explore the possible \textit{quantum utility} or \textit{usefulness} offered by various quantum computers for radio astronomy calibration. 

\section{Redundant Baseline Calibration and \texttt{heracal}}

We give here a very brief overview of the redundant calibration process and refer the reader to previous work~(\citep{noordam1982,wijnholds:2012,Dillon_2020}) for further explanation. Without loss of generality the visibilities observed between antennas $i$ and $j$, noted $V_{ij}^{\textnormal{obs}}(\nu, t)$, can be expressed as:

\begin{equation}
V_{ij}^{\textnormal{obs}}(\nu, t) =  g_i(\nu, t) g^*_j(\nu, t) V_{ij}^{\textnormal{true}} + \eta_{ij}(\nu, t)
\end{equation}

\noindent where $g_i$ is the gain of the i-th antenna, $\eta_{ij}$ the Gaussian thermal noise in the measurement and $V_{ij}^{\textnormal{true}}$ the true value of the visibility. Direction-independent calibration is a process for finding the solution of this system of equations such that it minimizes $\chi^2$ defined as:

\begin{equation}
    \chi^2  = \frac{1}{\textnormal{DoF}} \sum_{i<j}\frac{|V_{ij}^{\textnormal{obs}} - g_ig^*_j V_{ij}^{\textnormal{true}}|^2}{\sigma_{ij}^2}
    \label{eq:chisq}
\end{equation}

\noindent with $\textnormal{DoF}=N_{obs} - N_{ant} -N_{ubl} + 2$  is the number of degrees of freedom with $N_{ubl}$ the number of unique baseline. Note that we dropped the dependency on time and frequency for clarity. This process is usually done in multiple steps each bringing the solution to a lower value of $\chi^2$. The first step, referred to as \texttt{firstcal} exploits the fact that observed visibilities from the same redundancy group probe the same true visibility to solve for delay and overall phase factor in the gain. Writing the gains as:

\begin{equation}
%    g_j(\nu) = A_j(\nu)\exp \left( 2i\pi\nu\tau_j+i\theta_j \right)
    g_j(\nu) = A_j(\nu)~e^{( 2\mathrm{i}\pi\nu\tau_j+i\theta_j)} 
\end{equation}

\noindent where $A_j(\nu)$ is the amplitude of the signal, $\tau_j$ a delay corresponding to light travel time from the antenna to the correlator and $\theta_j$ an overall phase term to account for antenna feeds accidentally installed with a 180 degree rotation. We ignore the frequency dependence of the phase term here. Looking at two observed visibilities $V_{ij}$ and $V_{kl}$ that probe the same redundant baseline, $V_{\alpha}$, we can write:

\begin{eqnarray}
    \frac{V_{ij}V_{kl}^*}{|V_{ij}||V_{kl}|} &=& \frac{g_ig_j^*V_{\alpha}g_k^*g_lV_{\alpha}^*}{|g_i||g_j||g_k||g_l||V_\alpha|^2} \\
%    &=& \exp\big[i(\theta_i-\theta_j-\theta_k+\theta_l) \\ \nonumber 
%    && \quad \quad    +2i\pi\nu(\tau_i-\tau_j-\tau_k+\tau_l)\big] 
    &=& e^{[\mathrm{i}(\theta_i-\theta_j-\theta_k+\theta_l) + 2\mathrm{i}\pi\nu(\tau_i-\tau_j-\tau_k+\tau_l)]}
\end{eqnarray}

We make use of the redundant baselines in a regular array, giving this method its name. This set of equations is further linearized following the FFT based Quinn's second estimator. This leads to two large systems of equations, one for  delay and one for  phase. Following \citep{Gorthi_2020} we assess the redundant baseline calibration via the uncertainty in the antenna gains. This quantity is obtained by solving the linear system of equations multiple times, giving the antenna gains and computing the covariance matrix : $C = \big< \mathbf{g}\mathbf{g}^T\big>$. The diagonal of the covariance matrix gives the uncertainty per antenna, $\sigma_g$. 

The results of this process, referred to as \texttt{firstcal} in the Hera software, are  used as a \textit{good enough} starting point for a damped fixed-point iteration referred to as \texttt{omnical}\citep{Dillon_2020}. This approach iteratively improves the estimation of the gains and visibilities following:

\begin{equation}
    g_i^{n+1} = (1-\delta)g_i^n + \frac{\delta g_i^n}{\sum_j w^n_{ij}} \sum_j w_{ij} V_{ij}^{\textnormal{obs}}/y_{ij}^n 
\end{equation}

with $y_{ij}^n = g_i^ng_j^{n*}V_{\alpha}^n$ and $w^n_{ij} = (y_{ij}^n / \sigma_{ij})^2$ and $\delta$ a damping factor. During the iteration the estimation of the visibilities are also iteratively updated following a similar expression :

\begin{equation}
     V_{\alpha}^{n+1} = (1-\delta)V_{\alpha}^n + \frac{\delta V_{\alpha}^n}{\sum_{ij} w^n_{ij}} \sum_{ij} w_{ij} V_{ij}^{\textnormal{obs}}/y_{ij}^n
\end{equation}

The \texttt{firstcal} step relies on the solution of a linear system of equations to estimate the values of the gains. As such its time complexity scales with $\mathcal{O}(N_{ant}^3)$, i.e. the typical scaling for matrix inversion of singular value decomposition \citep{Dillon_2020}. The scaling can be improved to $\mathcal{O}(N_\textnormal{Ant} \log N_\textnormal{Ant})$ if one only considers the shortest baseline in the telescope at the cost of accuracy \citep{Gorthi_2020}.  In comparison, the time complexity of $\texttt{omnical}$ scales with $\mathcal{O}(N_{ant}^2)$ \citep{Dillon_2020} and is therefore much faster than $\texttt{firstcal}$ for large systems. In the following we therefore aim at replacing the classical solvers used in \texttt{firstcal} by quantum solvers and use the results of the quantum-enabled $\texttt{firstcal}$ step as a starting point for the $\texttt{omnical}$ fixed point iteration. We therefore only need to obtain a rough estimation of the gains using our quantum solvers and leave the refinement of these solutions to a classical fixed point iteration. We use in the following the \texttt{heracal} calibration pipeline \footnote{\url{https://github.com/hera_team/hera_cal}} and integrate the different quantum approaches directly into its linear solver module \footnote{\url{https://github.com/hera_team/linsolve}}

\section{Variational Quantum Linear Solver on Gate Based Quantum Computers}

\begin{figure}[h!]
    \centering
    \includegraphics[scale=0.35]{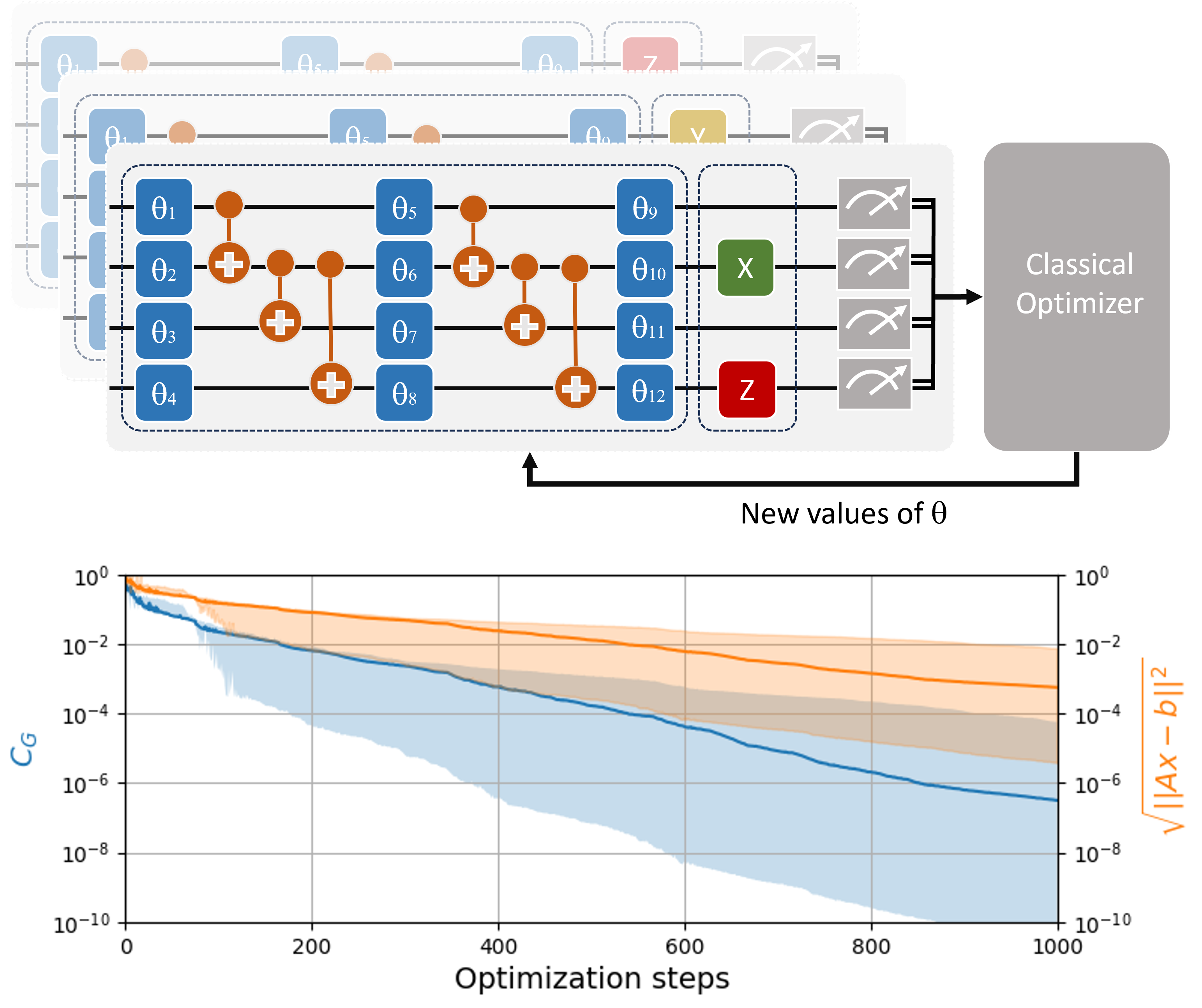}
    \caption{Top: Representation of the variational circuits used to solve linear systems following the VQLS approach. An variational ansatz with parameters $\theta_i$ is measured with different combination of Pauli gates. The results of these measurement is used classically to compute a cost function and optimize the values of $\theta_i$. Bottom: Evolution of the cost function during the 100 realization of the optimization process of the same linear system. The shaded area marks the variance over the ensemble of solutions while the bold line shows the mean value.}
    \label{fig:vqa}
\end{figure}

Variational quantum algorithms \citep{vqa_nature} have emerged as a attractive methods for harnessing the power of quantum computing in the NISQ era \citep{PreskillNISQ}. These methods rely on a so-called variatiational quantum circuit or ansatz, as the one depicted in Fig. \ref{fig:vqa} and that we denote here $V({\theta})$. These circuits are made up of a layer of single qubit rotation gates, whose rotation angles, ${\theta}$s, are the variational parameters, and entangling blocks made out of two-qubits CNOT gates. These layers are then repeated a number of times to allow the circuit to explore the full Hilbert space. As shown in Fig. \ref{fig:vqa}, several measurements are then performed on this circuit. The results of these measurements are then classically processed to compute a loss function and its gradients with respect to the variational parameters. New values of these parameters are calculated and encoded in the circuit before the process is repeated. \\

Variational approaches have been applied to many problems, such as the calculation of minimum energy state in quantum chemistry \citep{vqe_review} or solving of differential equations \citep{vqeels}. Recently the variational quantum linear solver (VQLS) approach \citep{vqls} has been proposed to solve linear system of equations using quantum circuits. In VQLS, the variational circuit is used to approximate the solution of the linear system, $x$, when applied to the $|0\rangle$ state: 

\begin{equation}
|x\rangle \approx V({\alpha})|0\rangle
\end{equation}

As such, the solution of the linear system is encoded in the amplitude of the quantum state created by the variational ansatz. This allows to efficiently use the Hilbert space of the system, as one only needs $\log_2(N)$ qubits to encode the solution of a $N\times N$ linear system. However accessing this solution is difficult and require the use of  quantum state tomography (QST) \citep{qst_qiskit} to fully characterize the amplitude and phase of the quantum state. Applying the matrix $A$ to the proposed solution $x$ leads to $|y\rangle = A|x\rangle = A V(\alpha) |0\rangle$. For this to work the matrix $A$ needs to be expressed as a sum of unitary matrices:

\begin{equation}
A = \sum_l^L c_l A_l
\label{eq:matrix_decomposition}
\end{equation}

To optimize the variational parameters, VQLS use a cost function based on the projection of $|y\rangle\langle y|$ on the subspace orthogonal to the rhs vector $|b\rangle$. The cost function reads :

\begin{equation}
C_G = \frac{\langle y | (\mathbb{I}-|b\rangle\langle b|) | y \rangle }{\langle y| y \rangle} = 1 - \frac{|\langle b| y \rangle|^2}{\langle y | y \rangle}
\label{eq:global_loss}
\end{equation}

The numerator and denominator of eq. (\ref{eq:global_loss}) can be written as :

\begin{equation}
\langle y | y \rangle = \sum_{ij} c_ic_j^*\bra{0}V^\dagger Aj^\dagger A_i V \ket{0}
\label{eq:hadamard_norm}
\end{equation}

\noindent and 

\begin{equation}
|\langle  b | y \rangle|^2 = \sum_{ij} c_ic_j^* \bra{0}U^\dagger A_i V\ket{0}\bra{0}V^\dagger A_j^\dagger U\ket{0} 
\label{eq:hadamard_overlap}
\end{equation}

\noindent where the operation $U$ creates the vector $|b\rangle$, i.e. the right hand side of the linear system from the $|0\rangle$ state: $|b\rangle = U |0\rangle$. The different terms in eqs (\ref{eq:hadamard_norm}) and (\ref{eq:hadamard_overlap}) can be evaluated through the Hadamard tests \citep{vqls}. The gradients of the loss function with respect to the variational parameters can also be computed using the same circuits or using the parameter-shift rule \citep{Wierichs2022generalparameter}. Using these gradients the values of the variational parameters can be iteratively updated to minimize the value of the loss function. A typical output of VQLS is represented in Fig. \ref{fig:vqa}. Starting from a relatively high value of the loss function, the progressive optimization of the rotation angles in the circuits reduces the value of cost function and by doing so reduces the residue norm of the linear system leading to a good approximation of the solution.

The quality of this approximation depends on the number of optimization steps and the optimizer used. It is also well known that when applying variational approaches to large systems,  barren plateaus, i.e. region where the gradients of the loss function vanishes, can significantly hamper the optimization process \citep{vqls}. Different variations of the cost function were proposed to limit the impact of these plateaus \citep{vqls}. We have implemented the VQLS approach in the Qiskit ecosystem and made our code publicly available \footnote{\url{https://github.com/QuantumApplicationLab/vqls-prototype}}.

\section{QUBO Linear Solver on Quantum Anneallers}

Quantum annealers (QA) are specialized quantum hardware that are very suited to solve optimization problems arising in many industrial cases \citep{QA_industry}. In QAs, the qubit system is made of interacting spins, where each spin can only take two values: $\sigma = \pm1$. The Hamiltonian of such system can therefore be written as:

\begin{equation}
H = \sum_n E_n\sigma_n + \sum_{nm} J_{nm}\sigma_n \sigma_m
\label{eq:ising}
\end{equation}

\noindent where $E_n$ is the energy of each qubit and $J_{nm}$ the interactions strength between two qubits. These parameters be carefully controlled as to favor parallel or anti parallel alignment between the different qubits. QAs solve problem by slowly turning on the interaction between the qubits and letting the system reach its ground state, where the total energy is minimized. 

% By encoding the variable in the spin values one can find the optimal combination of variables that minimizes the total energy. \\

Quadratic unconstrained boolean optimization problems (QUBO) seek to minimize a cost function written as:

\begin{equation}
E(q) = q^T M q = \sum_i M_{ii} q_i +\sum_{i,j} M_{ij} q_i q_j
\end{equation}

where each $q_i$ is a binary variable that can only take values $0$ or $1$. The similarities between this cost function and the underlying interactions between the qubits of a QA, eq. (\ref{eq:ising}) allows us to efficiently encode and solve QUBO problems on QAs \citep{ceselli_good_2023, Pastorello_2019}. \\

To solve a linear system using the QUBO formalism we first encode each  real-valued component, $x_i$, of the solution solution vector $x=\{x_0,x_1,...,x_N\}$ using a combination of binary variables $q_k$ \citep{FloatQuantAn}. Many different approaches have been developed to obtain an efficient binary encoding. In this paper we use the following encoding:

\begin{equation}
x_i = \frac{1}{2^{N_q} - 1}\left( -2^{N_q} q_{i,N_q} + \sum_{n=0}^{N_q-1} 2^{n} q_{i,n} \right)
\label{eq:encoding}
\end{equation}

\noindent where $N_q$ is the number of qubit used to express a single real number. The possible values of $x_i$ are here bound between -1 and 1 and the variable can take $2^N_q$ distinct discrete values in that interval.  \\

\begin{figure}[h!]
    \centering
    \includegraphics[scale=0.45]{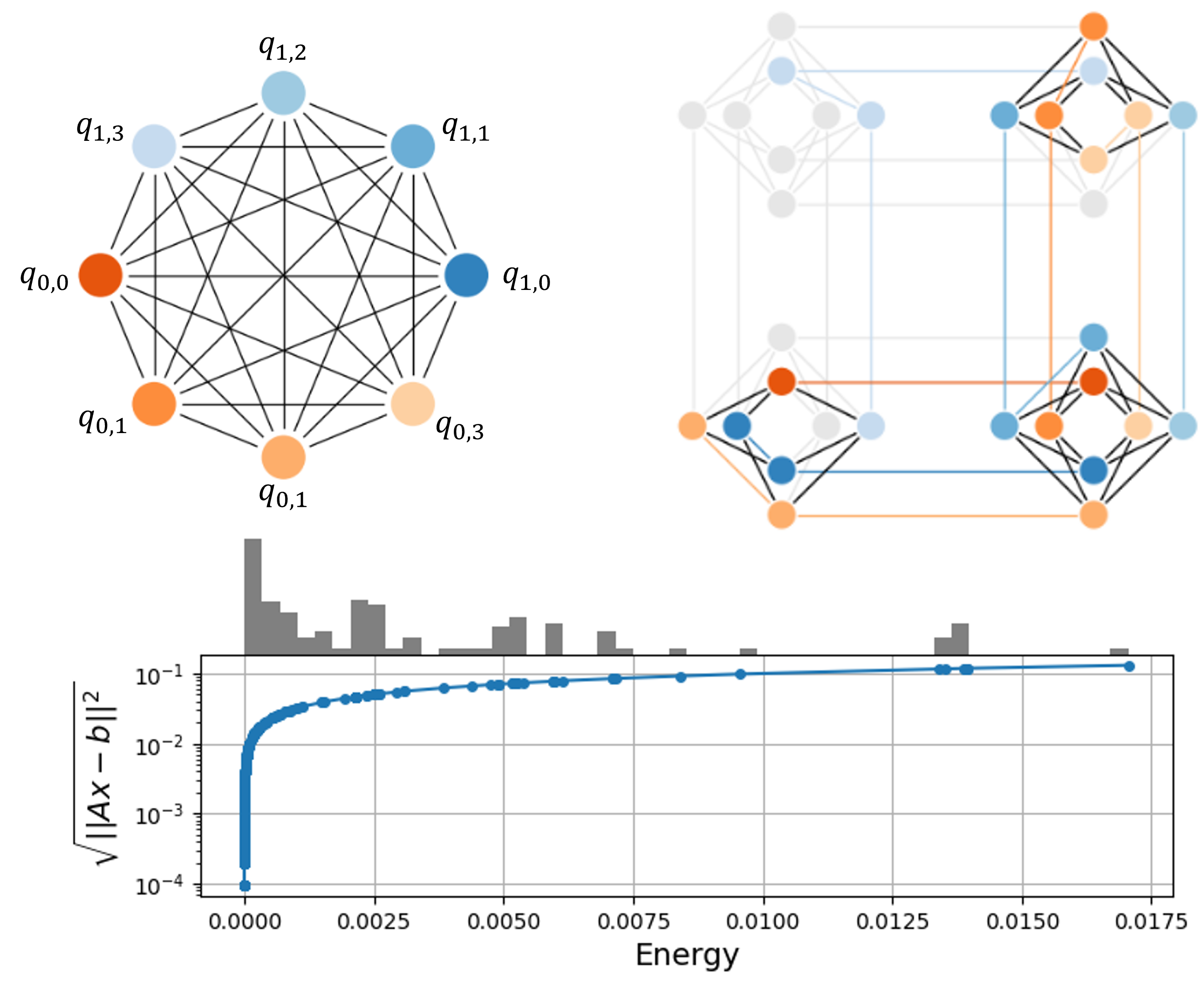}
    \caption{Top Left: Fully connected graph representing the QUBO formulation of a $2\times 2$ linear system with each variables encoded in 4 qubits: $q_{i,0},...q_{i,3}$. Top Right: Embedding of the fully connected graph into a \texttt{Chimera} connection graph of the D-Wave quantum computers. Each qubit $q_{i,j}$ is duplicated on several physical qubits to allow for all the necessary couplings. Qubits encoding for the same variables are show with the same color and are connected with a colored line. Bottom: Variation of the residue of a linear system with the energy of the QUBO formulation. The histogram on top of the plot indicates the final conformations obtained by the simulated annealing solver.}
    \label{fig:qubo}
\end{figure}

The cost function associated with solving the linear system can be directly defined through the norm of its residue vector \citep{ViableQUBOLS}:

\begin{eqnarray}
    E(x) &=& ||Ax-b||^2 = (Ax-b)^T(Ax-b) \\
    &=& x^T A^T A x - b^T A x - x^T A^T b +b^T b \\ 
    &=& x^T M x + ||b||^2
\label{eq:qubols_cost}
\end{eqnarray}

\noindent where $M$ is the QUBO matrix that encodes the interactions between the different qubits. The last term, i.e. the norm of the rhs vector, can be ignored as it does not contribute to the optimization process. Injecting the encoding given by eq. (\ref{eq:encoding}) in the cost function leads to the final QUBO formulation of the linear system. Fig. \ref{fig:qubo} illustrate the QUBO formulation of a $2\times 2$ linear system using 4 qubits per float in the encoding. As seen on this figure the QUBO matrix is fully connected, i.e. all the $q_{i,j}$ must interact with each other. Unfortunately, current QAs present a limited connectivity between their qubits. For example the \texttt{Chimera} connection graph of the D-Wave chip only allows for a maximum of 6 connections per qubits \citep{dwave_nextgen}. To embed the QUBO matrix on this chip it is therefore required to duplicate the logical variables $q_{i,j}$ onto several physical qubits to allow for all the necessary couplings. Physical qubit encoding the same logical variables are strongly coupled with each other, forming a chain, so that they ideally present the same spin values after the optimization process. The new \texttt{Pegasus} connection graph of next-generation D-Wave chip allow for 15 connections per qubits and therefore decreases the need for duplication of logical variables \citep{dwave_nextgen}. 

A typical result of this solver is shown in Fig. \ref{fig:qubo}. The QA is relaxed and sampled a number of times, each read leads to a particular configuration of the qubits spins. Each configuration has a specific energy that corresponds to a given value of the residue norm. Most read yield a low energy state and therefore a good approximation of the solution of the linear system. However, some reads reach a local minima with high energy and therefore a poor approximation of the solution. The read corresponding to the lowest energy is selected and the solution of the linear system is decoded from the corresponding qubit spin conformation. Note that at the difference of VQLS where quantum state tomography is required to extract the solution of the linear system from the ansatz, the solution of the linear system can here be simply reconstructed from the spin value of the qubits \citep{FloatQuantAn}. 

The viability of QAs for solving large scale linear systems has recently been studied in details \citep{ViableQUBOLS, QuantAnPolEq, FloatQuantAn}. One interesting recommendation emerging from these studies is to use QA to rapidly obtain an approximate solution and to further refine this solution using a classical fixed point iteration \citep{ViableQUBOLS}. This is precisely the strategy adopted by the redundant calibration pipeline, where the \texttt{firstcal} solutions are used as starting point of the \texttt{omnical} solver \citep{Dillon_2020}. We have implemented the QUBO linear solver in  our library \footnote{\url{https://github.com/QuantumApplicationLab/qalcore}} that can easily be reused for other purposes 

\section{Ideal Results}

We have integrated the VQLS and the QUBO solvers introduced above in the calibration pipeline for the HERA telescope ~\footnote{\url{https://github.com/QuantumRadioAstronomy/hera_cal}}. This software suite contains all the tooling for the operation, maintenance and analysis of the HERA telescope. The software is actively developed and used by the team of astronomers operating the telescope. The tight integration of the quantum solvers in this software suite enables to seamlessly switch between classical and quantum resources for the calibration and therefore easily explore the potential use of quantum resources for radio astronomy.   

In this section we consider the ideal case where the quantum computers used have no restriction in terms of qubit connectivity, coherence time and noise. We compare the results obtained with the classical and quantum solvers in terms of accuracy but not timing. Due to the computational cost required to simulate quantum computers, we restrict ourselves to relatively small hexagonal arrays containing 7, 19 and 37 antennas loosely based on real HERA layout. Following \citep{Gorthi_2020},  the calibration data were simulated by generating a set of true visibility for all unique baseline types and a set of gains for all antennas. The visibilities have a constant average over all baseline types. The gains are Gaussian distributed with an average amplitude of 1 and the gain scatter of 0.1. A Gaussian distributed noise of variable strength was added to the signal to simulate actual operating conditions. 

Unless otherwise specified, the VQLS solver used a real-amplitude variational ansatz with 3 repetitions and a full correlation between the qubits. As the name indicates, a real-amplitude ansatz ensures that quantum state generated by the ansatz only contains real-valued amplitudes. This restriction does not affect our results as the solution of the linear system considered here are all real. The optimization process uses the cost function given in eq. (\ref{eq:global_loss}) with 500 optimization steps of the gradient-free optimizer COBYLA \citep{cobyla_vqe, cobyla_vqls}. Using a different cost function less affected by barren plateaus with more iterations and a more accurate optimizer may improve the accuracy of the final results. However, we show below that the rough optimization strategy adopted here leads to sufficiently accurate results. Finally, we assume here that one can have directly access to the full quantum state of the circuit to read the final solution of the linear solver. In other word we do not have to sample the circuit wave function and perform quantum state tomography to extract the solution. 

For the QUBO solver, we used 11 qubits to encode each floating point number of the solution vector, which results in an accuracy of about $10^{-3}$ in the encoding. While exact sampler could be used to determine the exact ground state energy of the system given the small number of variables, we use a simulated annealing sampler here that more accurately simulates the performance of real QAs. For each linear system, we used 1000 independent minimizations of the cost function to reach the global minimum of the problem. In this experiment we assume that each qubit is connected to all the other qubits so we do not have to embed our QUBO problem in a graph with limited connectivity. 

\begin{figure}[h!]
\begin{center}
\begin{tabular}{c}
\includegraphics[scale=0.5]{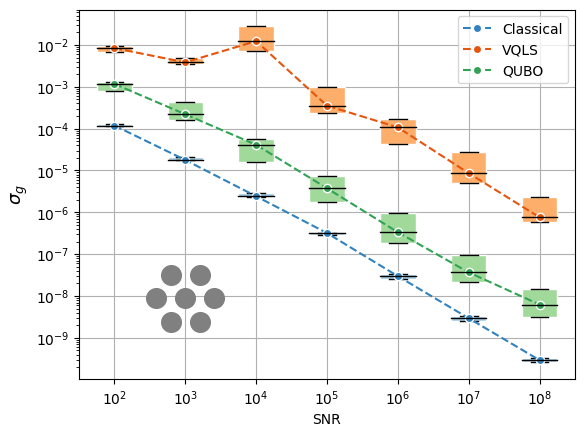}\\
\includegraphics[scale=0.5]{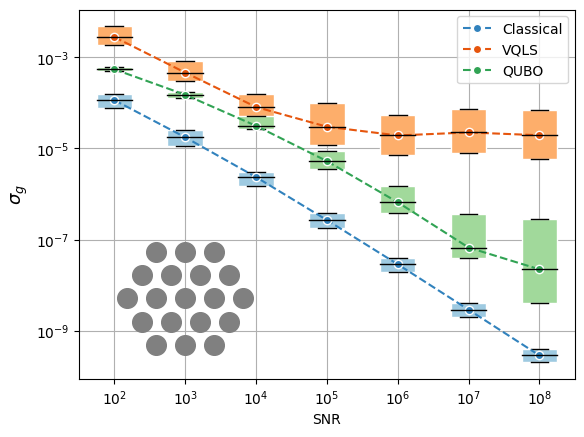} \\
\includegraphics[scale=0.5]{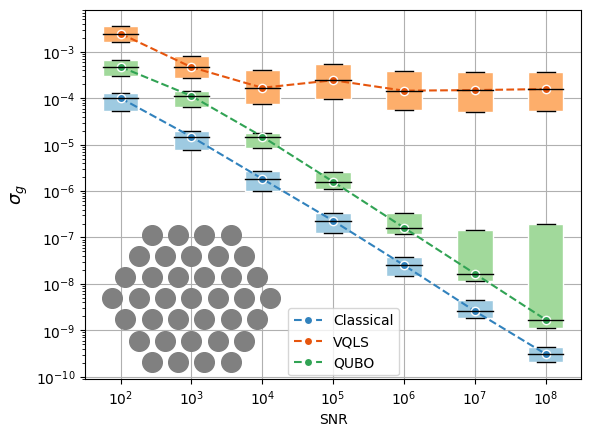}
\end{tabular}
\caption{\label{fig:comparison_firstcal} Comparison between the gain variance obtained for hexagonal arrays after the \texttt{firstcal} step for the different classical and quantum methods. }
\end{center}
\end{figure}

Fig. \ref{fig:comparison_firstcal} shows the results of the \texttt{firstcal} calibration for three different arrays, using the classical approach as well as the two quantum methods considered here. This figure shows the variance of the gain covariance with respect to the signal to noise (SNR) ratio given by the noise intensity injected in the system. We explore a wide range of SNR values to illustrate the shortcomings of the two quantum solvers. In each case the median value of the gain covariance over 100 independent solutions is represented as well as the 25\% and 75\% quartile. A low covariance gain indicates a good calibration where the gain values do not fluctuates largely over the array. \

The gain variance obtained with the classical approach leads to a well known $\textnormal{SNR}^{-2}$ scaling. Classical solvers are therefore capable to reach accurate calibration even with very low SNR. The QUBO solver shows a similar behavior, albeit with higher $\sigma_g$ values. The difference between the classical solver and the QUBO solver comes from the floating point encoding of eq. (\ref{eq:encoding}). Due to this encoding, even the lowest energy state does not give exactly the solution of the linear system as it is limited by the maximum floating point precision attainable. The effect of this limited precision is even more noticeable for high SNR values leading to a large variance of $\sigma_g$.  Nevertheless the calibration obtained with the QUBO solver is comparable with the classical one over the whole range of SNR values explored here. 

The VQLS solver leads to higher values of $\sigma_g$ and fails to reproduce the $\textnormal{SNR}^{-2}$ scaling for the larger arrays. This is largely due to the rough optimization strategy we have adopted here. The fixed small number of optimization steps, combined with the gradient-free optimizer, leads to an incomplete optimization process not able to reach a very accurate approximation of the solution. As seen in the top picture this optimization can also be unstable leading here to unexpectedly high values of $\sigma_g$ for $\textnormal{SNR}=10^4$. In addition, as the system size grows, the optimization because more difficult as the number of parameters in the variational ansatz increases. This leads to a plateauing of the $\sigma_g$ values for the larger arrays. 

To assess the suitability of the different solvers in the calibration pipeline we have computed the values of $\chi^2$ via eq. (\ref{eq:chisq}) after the \texttt{firstcal} calibration step for the three different arrays. Since the true redundant visibilities $V_\alpha$ are not solved for in \texttt{firstcal}, their values are approximated by the mean value of the observed visibilites in their redundant group. 

\begin{figure}[h!]
    \centering
    \begin{tabular}{c}
    \includegraphics[scale=0.5]{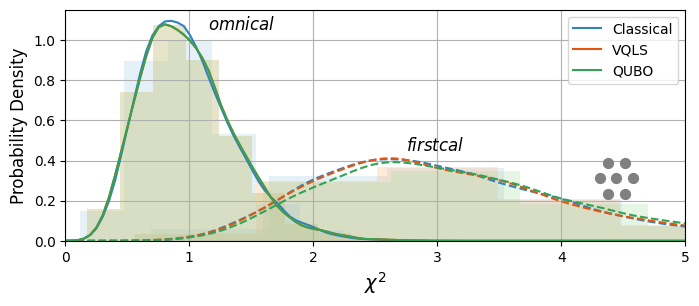} \\
    \includegraphics[scale=0.5]{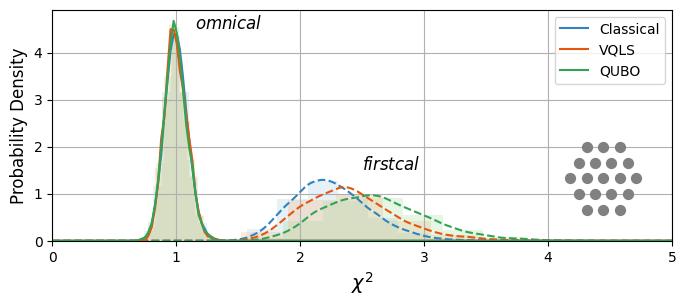} \\
    \includegraphics[scale=0.5]{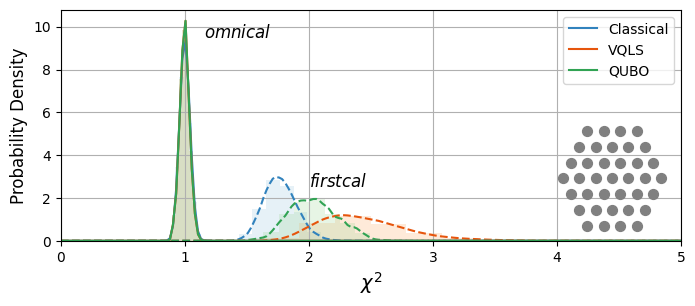} \\
    \end{tabular}
    \caption{Value of $\chi^2$ obtained after \texttt{firstcal} (dashed line) for the different methods considered here.  The plain lines shows the distribution of $\chi^2$ values obtained after the $\texttt{omnical}$ step starting from the corresponding \texttt{firstcal} results. Results obtained with $\textnormal{SNR}=10^2$} 
    \label{fig:chisq}
\end{figure}

Fig. \ref{fig:chisq} shows the values of $\chi^2$ obtained for $\textnormal{SNR}=10^2$. As seen on this figure, the distribution of $\chi^2$ after \texttt{firstcal} differs only slightly when using the classical a quantum solvers for the two smaller arrays. The difference is more apparent for the larges array where, as expected from the results presented in Fig. \ref{fig:comparison_firstcal}, VQLS tends to perform the worst of the three methods. Further comparison between the accuracy of the solutions provided by the quantum solver is provided in Fig. \ref{fig:residues}. As seen in this figure, the solutions obtained via the quantum solvers deviate from the exact solution due to the limitation of the optimization processes chosen for both methods.

We have then used the estimation of the gain values obtained after \texttt{firstcal} with the three methods as starting point of the \texttt{omnical} fixed-point solver. Here, only a good enough estimation of the gains is necessary for \texttt{omnical} to converge in a few iterations. As seen in Fig. \ref{fig:chisq}, distribution of $\chi^2$  obtained with the three different methods after \texttt{omnical} are nearly identical. This demonstrates that, despite their limitations, the quantum solvers yield sufficiently accurate estimations of the gains that can be reliably used as starting point for \texttt{omnical}. 

\begin{figure}
    \centering
    \begin{tabular}{c}
        \includegraphics[scale=0.45]{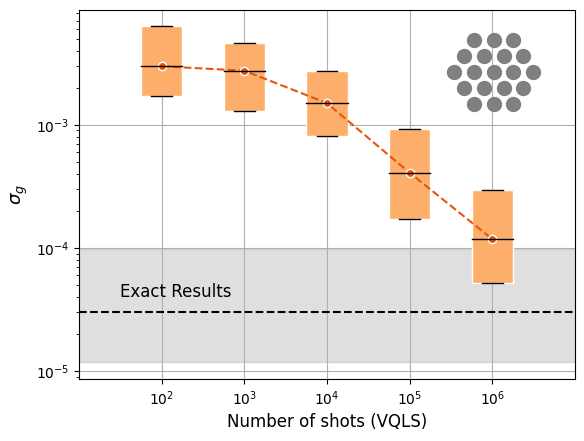} \\
        \includegraphics[scale=0.45]{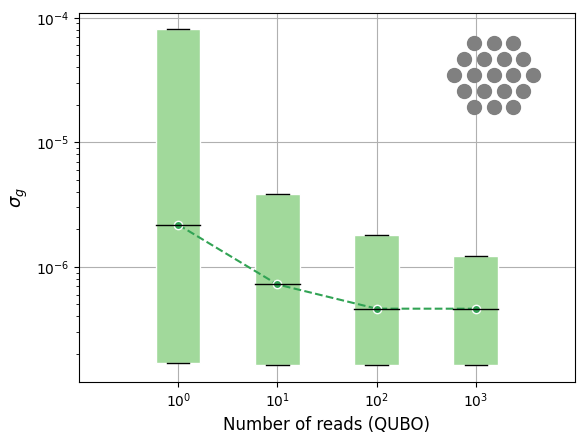}
    \end{tabular}
    \caption{Performance of the VQLS and QUBO linear solver with the number of shots used to sample the circuit output and number of reads of the annealer results respectively. The results were obtained for a 19 antennas hexagonal array with $\textnormal{SNR}=10^{-6}$}
    \label{fig:shots}
\end{figure}

We finally explore how some parameters of the quantum computers might affect the accuracy of the linear solver solution. For the VQLS solver, we have so far assumed that the cost function was computed with a perfect knowledge of the quantum state of the circuit. In practice this is not the case and we have to sample the outcome of the different measurements to assemble the cost function via so-called so-called shots, i.e. independent preparation and measurement of the quantum circuits. Fig. \ref{fig:shots} shows the evolution of $\sigma_g$ for a 19 antennas array with $\textnormal{SNR}=10^{-6}$ with respect to the number of shots used to sample the quantum state and evaluate the cost function. As seen in this figure, while the values of $\sigma_g$ are not too bad for number of shots as low as 100, more than $10^6$ shots would be necessary to recover the accuracy obtained assuming that the quantum state of the system is readily and precisely available. This is an important limitation as the computational cost increases linearly with the number of shots. In addition the number of shots possible on real quantum hardware is often limited, for example to 300K shots on the IBM quantum chips currently available. 

The accuracy of the results obtained with the QUBO solver depends on the number of time we solve the system, i.e. the so-called number of reads. Fig. \ref{fig:shots} shows the evolution of $\sigma_g$ for the same system described above, as a function of the number of reads. As seen in this figure, even for a single read, the accuracy of the solution is very high, albeit with a large variance since this single read might reach a high energy state. However increasing the number of reads to 10 or 100 is sufficient to recover the accuracy obtained with 1000 reads in Fig. \ref{fig:comparison_firstcal}. This indicate that only a very small number of reads might be sufficient if \texttt{firstcal} only needs to output a rough estimation of the solution.  

\section{Results with realistic settings}

In contrast to the idealized situation explored in the previous section, current quantum computers have significant limitations. While impressive advances are being made to improve the specifications of quantum chips in terms of qubit connectivity and coherence time, the quantum computers available at the moment are still quite limited. 

Quantum gate based computers, such as the IBMQ machines, have coherence time in the order of 100 ms. While this is an impressive achievement, it limits the depth of the circuit one can accurately simulate. Noise mitigation techniques are being actively developed \citep{ZNE} and have been show to lead to impressive results when applied to real world problems \citep{utility}. In addition, qubit connectivity is limited to the nearest neighbors of each qubit, which limits the entanglement map we can use in the variational ansatz. This limited connectivity also forces the introduction of ancillary qubits when a higher degree of connectivity is needed, for example in the Hadamard tests required to compute the VQLS cost function. 

QAs such as the D-Wave chips have limited connectivity graphs for each qubit. For example, \texttt{Chimera} topology only support a maximum of 6 connections per qubit. This is far from being enough to encode the QUBO problems studied here, which requires full connectivity between the qubits. To remediate this issue, several qubits must be tightly coupled together to form a chain that encodes the same variable while increasing the number of potential coupling to other variables. This chains can however break during the optimization, i.e. qubits in the chain end up with different state, reducing the accuracy of the solution. Furthermore, this requires many more qubits than would be required in a fully connected QA.

\begin{figure}[h!]
    \centering
    
\begin{tabular}{c}
    \includegraphics[scale=0.5]{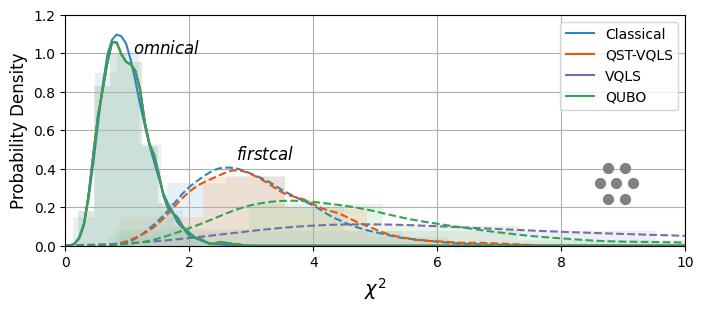} \\
    \includegraphics[scale=0.5]{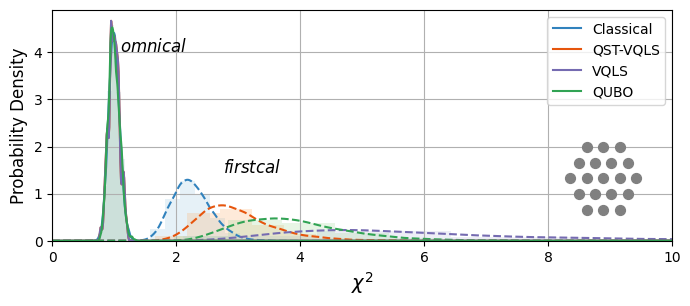} \\
    \end{tabular}
    \caption{Value of $\chi^2$ obtained after \texttt{firstcal} (dashed line) for the different methods considered here using realistic simulation of noisy quantum computers.  The plain lines shows the distribution of $\chi^2$ values obtained after the $\texttt{omnical}$ step starting from the corresponding \texttt{firstcal} results. The results were obtained for a 7 and 19 antennas hexagonal array with $\textnormal{SNR}=10^2$}
    \label{fig:chisq_noisy}
\end{figure}

Fig. \ref{fig:chisq_noisy} shows the distribution of $\chi^2$ obtained from simulations performed with realistic settings for the quantum computing hardware. As explained in details below, the VQLS results were obtained with realistic noise models of IBMQ quantum computers, while the QUBO results account for the limited connectivity offered by D-Wave chips. The simulations were performed only for the smaller two arrays due to the large computational cost involved. 

\subsection{VQLS on IBM-Q Gualdalupe}

To emulate the behavior of the real quantum hardware during the VQLS optimization, we have made use of the \texttt{Aer} simulators \citep{qiskit} provided by the Qiskit development tool suite. These simulators are capable of reproducing the expected behavior of real quantum circuits by accounting for their inherent relaxation and decoherence processes. We have used the noise model of the IBM Guadalupe, a Falcon 16 qubits chip. 

As explained in \ref{Appendix:QST-VQLS} and illustrated in Fig. \ref{fig:udepth}, one the main limitation of VQLS when executed on real quantum hardware comes from the circuit $U$, required to create $|b\rangle$ and used in eq. \ref{eq:hadamard_overlap} \citep{read_fine}. As show in Fig. \ref{fig:udepth}, the depth of this circuit increase exponentially with the system size and any measurement accounting for realistic noise mode of any circuit containing $U$ is unusable due to noise (see Fig. \ref{fig:udepth}). Similarly, any matrix decomposition, eq. (\ref{eq:matrix_decomposition}), that leads to deep circuit implementation of the different unitary terms introduce a significant amount of noise in the circuit measurement.

To circumvent this issues and enable the execution of VQLS on noisy simulators, we propose here a variation of the original VQLS algorithm  briefly presented in \ref{Appendix:QST-VQLS}. This approach, coined Quantum State Tomography Variational Linear System (QST-VQLS), exploits fast calculation of the full density matrix of the variational ansatz through classical shadows \citep{classical_shadows, shadow_noise, adapt_shadow}, to estimate the different terms occurring in the loss function. In addition a simple Pauli decomposition of the matrix $A$ is here adopted to limit the depth of the circuits. This decomposition however leads to a very large number of terms in eq. (\ref{eq:pauli_decomposition}). It is important to note that while QST-VQLS allows to improve the accuracy of VQLS in presence of realistic noise sources, its computational cost is significant as it requires to perform an accurate QST at each optimization step. 

As seen in Fig. \ref{fig:chisq_noisy}, the performance of the original VQLS algorithm is rather unsatisfying due to the depth of the circuits encoding $U$ and the different $A_l$. This leads to a very broad distribution of $\chi^2$. In contrast the results obtained with the QST-VQLS approach are very close to the ones obtained with the classical solver. The performance improvement offered by QST-VQLS is also visible from Fig. \ref{fig:residues} where its solution are visibly much more accurate than the ones obtained with VQLS. 

While these results seem encouraging it is important to note that no quantum speed up can be obtained with the QST-VQLS approach. One of the leading term in the computational complexity of the VQLS approach comes from the double sum in eq. (\ref{eq:hadamard_overlap}). This sum leads to $L(L+1)/2$ terms where $L$ is the number if terms in the matrix decomposition eq. \ref{eq:symmetrix_decomposition}. This sum must be evaluated at each optimization step, leading to a scaling of $N_{opt} L(L+1)/2$ classical operations needed to solve the linear system. Fig. \ref{fig:scaling_final} shows this scaling as a function of the number of antennas in the system for different values of $L$. The Pauli decomposition, eq. \ref{eq:pauli_decomposition} used in the QST-VQLS simulations leads to $L\approx N_{ant}^2$ terms in the decomposition. This leads to an overall computational cost far larger than the classical methods scaling with $N_{ant}^4$. The symmetric decomposition of eq. (\ref{eq:symmetrix_decomposition}) leads to $L=2$. However, as mentioned above the circuits require to encode the resulting unitary matrices are too deep for current architectures. The cost of calculating the decomposition and creating the corresponding circuits is also significant and therefore the symmetric decomposition does not provide a viable solution.

As seen in Fig. \ref{fig:scaling_final}, the HERA telescope will contain 320 antennas. At this scale a sub-linear scaling: $L<N_{ant}$ will be required for VQLS to potentially be competitive with classical approaches. Even when extrapolating our results to the scale of the SKA-LOW telescope that will contain more that 100000 antennas, an efficient matrix decomposition is needed for VQLS to provide any advantage compared to classical approaches. 

\subsection{QUBO on \texttt{Chimera} topology}

\begin{figure}
    \centering
    \includegraphics[scale=0.5]{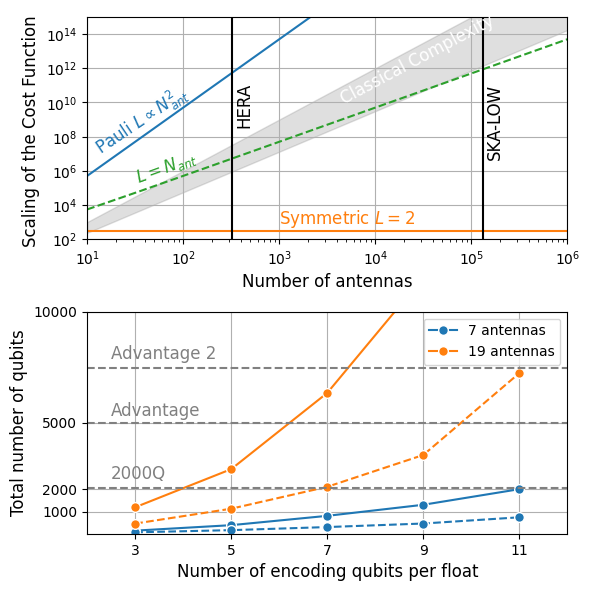}
    \caption{Top: Scaling of the VQLS cost function with the number of antennas in the system for different value of $L$, i.e. the number of terms in the decomposition of the matrix $A$. Only for decomposition scaling sub-linearly with the number of antennas can a speed up potentially be obtained. Bottom: Total number of required qubits for arrays containing 7 and 19 antennas with respect to the number of qubits used in the encoding of the QUBO problem. The full(dashed) line corresponds to an embedding on the \texttt{Chimera}(\texttt{Pegasus}) graph with 6(15) connection per qubit.}
    \label{fig:scaling_final}
\end{figure}

To account for the limited connectivity offered by D-Wave chip, we have embedded the QUBO systems of our different antenna arrays into a \texttt{Chimera} graph, used by most of the D-wave platform, containing $64\times64$ unit cells. Such embedding is a critical task for the application of QAs to real world problems and major advancements are currently being made to improve its performance \citep{chimaira_embedding, chimaira_embedding_2, chimaira_embedding_3, chimaira_embedding_4}. With each unit cell containing 8 qubits, this leads to a 32768 qubits. While this could seem excessive to encode for a maximum of $37\times11=333$ variables, Fig. \ref{fig:scaling_final} shows that the number of qubits required to encode a fully connected QUBO problem of 19 floating point variables encoded in 11 qubits each already exceeds 20000 on the \texttt{Chimera} graph. Embedding the QUBO problem on the \texttt{Pegasus} graph reduces the number of required qubits due to its increased connectivity. Fig. \ref{fig:scaling_final} shows that the current available D-Wave chips, 2000Q, Advantage and Advantage 2, with respectively 2048, 5000 and 7500 qubits, can only accommodate the two smaller arrays when using 11 encoding qubits per float. 

Such embedding leads to very long qubits chains, and chain breaking is frequent even for small array sizes. While such chain breaking is mitigated by performing majority voting to restore the consistency of the encoding, they result in a significant performance decrease for the solution of the linear systems. Consequently, the distribution of $\chi^2$ obtained after \texttt{firstcal} with the realistic QUBO solvers are far worse than was shown previously for the ideal case. In addition, Fig. \ref{fig:residues} clearly shows that the solutions provided by the QUBO solver deteriorate greatly when accounting for the limited connectivity of the quantum chip.

\section{Conclusion}

As quantum computers become increasingly available, it is important to assess the possibilities this emergent technology can offer to the scientific community. In this paper we explored how quantum linear solvers can be used for the calibration pipeline of large-scale radio telescopes. To this end we have integrated a variational and a combinatorial quantum linear solver into the redundancy calibration pipeline of the HERA telescope. Our results show that, in the ideal case where the different quantum hardware are not limited by the qubit connectivity or the coherence time, quantum linear solvers are a viable tool to obtain the initial estimation of antennas gains. However, current limitations in terms of qubit connectivity and coherence time significantly hamper the performance of the investigated quantum linear solvers. While the variational approach implemented on gate-based quantum computers requires a relatively small number of qubits, even for large arrays, it requires a very efficient matrix decomposition scheme to potentially compete with classical methods. Without such a decomposition its computational cost is prohibitive. The combinatorial approach, that relies on quantum annealers, yields very accurate results but require a very large number of physical qubits to compensate for the limited inter-qubit connectivity on real devices. Consequently, current quantum annealers can only accommodate small antenna arrays where classical approaches are fast enough. 

We therefore conclude that, at least with the hardware currently available, no tangible quantum advantage can be identified. 
However, we have shown that, under ideal conditions, both investigated quantum linear solvers are viable alternatives to the computationally expensive initial loop of redundancy calibration loop. 
Any  advantage is negated by limitations in current hardware. 
Therefore, exploration of the opportunities offered by quantum computers are necessary to properly assess their potential benefit for radio astronomy. 
New quantum solvers with better performance and new hardware with less stringent limitations will emerge that are likely to bring computational advantages. 
The computational pipeline described in this paper can easily be extended to include new development and allow for a rapid evaluation of their benefits.

\section*{Acklowledgements}
This work was done by the Netherlands eScience Center in partnership with ASTRON under project number 27020P01. We extend our thanks to Dr. Stefan Wijnholds (ASTRON) for their helpful discussions and review of an early version of this manuscript. The authors would also like to thank Joshua Dillon for providing help in the utilization of the hera\_cal software. This paper is part of a broader investigation into the feasibility of quantum computers in radio astronomy. We thank Thomas Brunet, Dr. Emma Tolley and Dr. Stefano Corda (EPFL) and Roman Ilic and Dr. Daniel C\'{a}mpora P\'{e}rez (University of Maastricht) for the interesting discussions and refer to their accompanying paper in this same journal.

\bibliographystyle{elsarticle-harv}
\bibliography{references}

\appendix

\section{Symmetric Matrix Decomposition}
\label{appendix:symmdec}
\begin{figure}[h!]
    \centering
    \includegraphics[scale=0.5]{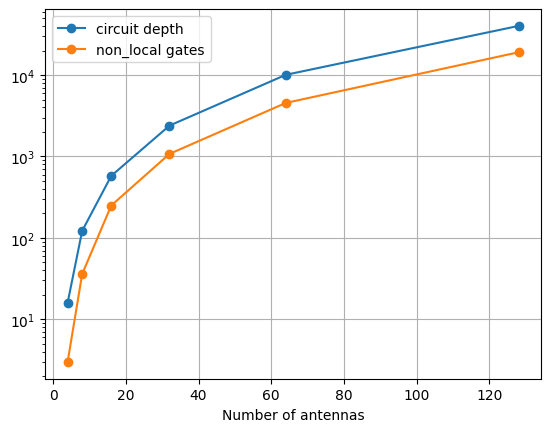}
    \caption{Depth and number of non-local gates required to transpile the $A_\pm$ matrices in quantum circuits for different number of antennas in a linear array.}
    \label{fig:adepth}
\end{figure}

Any real symmetric matrix $A$ with norm 1, can be decomposed in two unitary matrices following $A = A_+ + A_-$ with:

\begin{equation}
A_\pm = \frac{1}{2}(A \pm i\sqrt{I-A^2})
\label{eq:symmetrix_decomposition}
\end{equation}

\noindent after normalization of $A$. The two matrices $A_{\pm}$ are both unitary and can therefore be used to encode $A$ in a quantum circuit. However, computing and transpiling these matrices into a quantum circuits is computationally demanding. In addition of the classical computational cost, the depth of the corresponding quantum circuits and the number of non-local gates increase rapidly with the matrix size  as seen in Fig. \ref{fig:adepth}

\section{Quantum State Tomography Variational Quantum Linear Solver with Pauli Matrix Decomposition}
\label{Appendix:QST-VQLS}

We briefly present here the QST-VQLS with Pauli decomposition approach used to execute VQLS on realistic quantum architectures. 

\paragraph{Pauli Decomposition: } VQLS requires to find a decomposition of the matrix $A$ in a series of unitary matrices (see eq. \ref{eq:pauli_decomposition}). We consider here the case where the matrix $A$ is decomposed as a series of Pauli matrices:

\begin{equation}
A = \frac{1}{N_P} \sum_{i,j...,N_P} a_{ij..N} \sigma_i \otimes \sigma_j \otimes ... \otimes \sigma_{N_q}
\label{eq:pauli_decomposition}
\end{equation}

\noindent where $\sigma_k \in {I, X, Y, Z}$ is a 1-qubit Pauli gate or Identity, $N_P$ the total number of Pauli strings in the decomposition and $N_q$ the number of qubits used. The coefficients of the sum are simply given by: 

\begin{equation}
  a_{ij..N_q} = \textnormal{Tr}(\sigma_i \otimes \sigma_j \otimes ... \otimes \sigma_{N_q} A)  
\end{equation}
 
Pauli decomposition lead to a large number of terms that scale with: $N_P \propto N_{\textnormal{Ant}}^2 $. This is a well known bottleneck of quantum computing approach that also limits its applicability to other domains. For example a VQE used to compute the ground energy of a molecular system,  requires the evaluation of $N_\textnormal{elec}^4$ circuits where $N_\textnormal{elec}$ is the number of electrons in the system.

We show in the following how to alleviate the computational requirements for the calculation of the different terms of eq. (\ref{eq:global_loss}).

\subsection{Evaluation of $\langle y|y \rangle$: Pauli Contraction \& Measurement Optimization}

\begin{figure}[ht!]
\begin{center}
\includegraphics[scale=0.55]{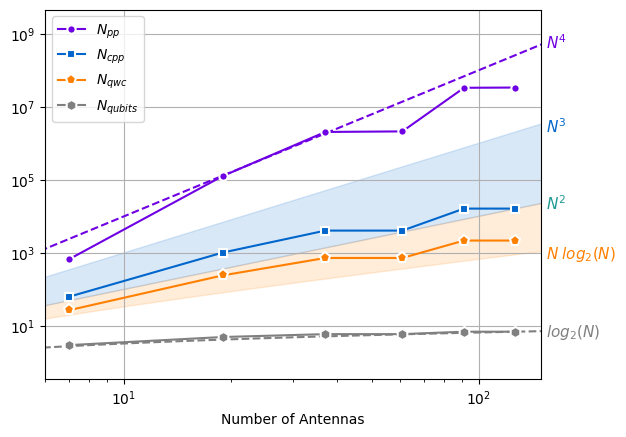}
\caption{\label{fig:num_circuits} Scaling of the main quantities involved in the calculation of $\langle y | y\rangle$. The number of qubits scale with $log N$. The number of pauli pairs initially scaled with $N^4$ but can be brought down to $N^2$ using Pauli contraction and to almost $N log N$ using measurement optimization techniques.}
\end{center}
\end{figure}

The calculation of $\langle y|y \rangle$ through eq. (\ref{eq:hadamard_norm}) using a Pauli decomposition of $A$ would require a number of Pauli pairs $A_iA_j^\dagger$,  scaling as $N_{pp}\propto N_{\textnormal{Ant}}^4$. As seen in Fig. \ref{fig:num_circuits} this would require the evaluation of millions of Hadamard Tests even for moderate system size. Given the current scarcity of quantum resources this direct approach is out of reach. 

\paragraph{Pauli Contractions: } The number of required quantum circuits can be significantly reduced. The first step to do so is to simplify the Pauli pairs $A_j^\dagger A_i$ in a single contracted Pauli string. 

\begin{equation}
 A_j^\dagger A_i = \alpha_{ij\rightarrow k} A_{k}   
\end{equation}

\noindent where ${ij\rightarrow k}$ denote the mapping of the $i,j$ combination onto the  Pauli string $A_k$. These coefficients can easily be derived from the table \ref{table:pauli_contraction}. Many combinations $A_j^\dagger A_i$ are then equivalent to the same $A_k$ which allows to write $\langle y | y \rangle$ as :

\begin{equation}
\langle y | y \rangle = \sum_{k}^{N_{cpp}} \alpha_k  \bra{0}V^\dagger A_k V \ket{0}
\label{eq:hadamard_norm_contracted}
\end{equation}

\noindent with $\alpha_k = \sum_{i,j\in K_{ij}} c_ic_j^* \alpha_{ij\rightarrow k}$ where $K_{ij}$ denote all the Pauli pairs $(i,j)$ that can be contracted to the $K$-th Pauli string. When applied to our calibration problem the number of contracted Pauli strings scale to $N_{cpp} \propto N_\textnormal{Ant}^2$ (Fig. \ref{fig:num_circuits}). 

In addition to reducing the number of quantum circuits that needs to be evaluated, the contraction of the Pauli pairs also allows us to replace the Hadamard test used to evaluate the components of $\langle y | y \rangle$ by direct measurements of the expected values of the $A_k$ operator \citep{Mitarai_2019}. As illustrated in Fig. \ref{table:hadamard}, this removes entirely the need to control the $A_l$ sub-circuits. This controlled operations would require the control qubit to be connected to all the other qubits in the circuits and therefore impose the use of ancillas qubits to compensate for the limited connectivity of real quantum chips.

\begin{table}[t]
\[
\begin{array}{c}
\Qcircuit @C=1.em @R=.7em {
\lstick{\ket{0}} & \qw                         & \gate{H}  & \ctrl{1}                          & \ctrl{1}                             & \gate{H} & \meter \\ 
\lstick{\ket{0}} & \multigate{2}{\mathcal{V}}  & \qw       & \multigate{2}{\mathcal{A}_i}      & \multigate{2}{\mathcal{A}_j^\dagger} & \qw      & \qw\\
\lstick{\ket{0}} & \ghost{\mathcal{V}}         & \qw       & \ghost{\mathcal{\mathcal{A}_i}}   & \ghost{\mathcal{A}_j^\dagger}        & \qw      & \qw \\
\lstick{\ket{0}} & \ghost{\mathcal{V}}         & \qw       & \ghost{\mathcal{\mathcal{A}_i}}   & \ghost{\mathcal{A}_j^\dagger}        & \qw      & \qw \\
}
\end{array}
\]
\vspace*{0.5cm}
\[
\begin{array}{c}
\Qcircuit @C=1.em @R=.7em {
\lstick{\ket{0}} & \multigate{2}{\mathcal{V}}         & \multigate{2}{\mathcal{A}_\alpha}     & \meter \\
\lstick{\ket{0}} & \ghost{\mathcal{V}}                & \ghost{\mathcal{\mathcal{A}_\alpha}}  & \meter \\
\lstick{\ket{0}} & \ghost{\mathcal{V}}                & \ghost{\mathcal{\mathcal{A}_\alpha}}  & \meter \\
}
\end{array}
\]
\caption{Circuits used to compute the different terms in the VQLS cost function: (top) $\bra{0}V^\dagger A_j^\dagger A_i V \ket{0}$ using the Hadamard test proposed originally (top) and the direct measurement used here (bottom)}
\label{table:hadamard}
\end{table}

\paragraph{Measurement Optimization: }The number of terms to be measured can be even further reduced using measurement optimization techniques \citep{gokhale2019minimizing, meas_opt}. These techniques take advantage of the fact that many operators $A_k$ in eq. \ref{eq:hadamard_norm_contracted} commutes with each other and can therefore be measured simultaneously in their shared eigen-basis. The identification of commuting operator is usually done via a qubit-wise commutation (QWC) approach but full commutation is also being explored. When applied to redundant calibration this leads to a number of terms scaling as  $N_\textnormal{Ant}^2 \leq N_{qwc} \leq N_\textnormal{Ant} \;log N_\textnormal{Ant}$. In generally  a maximum clique problem need to be solved to determine which operators can be measured simultaneously,  but new methods are being developed to alleviate this computational cost \citep{fastPauliGrouping}. 

 To circumvent this issue we here simply pick the first contracted Pauli pair in our list that does not contain the identity and identify all its QWC strings in the list. These strings form our first clique. We then repeat the process with the next contracted Pauli pair that does not contain the identity and that is not yet in a clique. This approach might not lead to the optimal clique identification but has a very small computational cost. 

\begin{figure}
    \centering
    \includegraphics[scale=0.45]{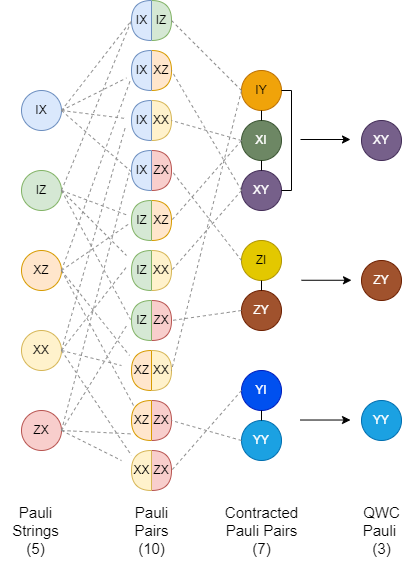}
    \caption{Illustration of the process allowing to reduce the number of circuits needed to compute $\langle y|y \rangle = \bra{0}V^\dagger A_j^\dagger A_i V \ket{0}$. }
    \label{fig:pauli_simplification}
\end{figure}

\paragraph{Example: }Let's illustrate the process of Pauli contraction and measurement optimization with a simple example. Figure \ref{fig:pauli_simplification} illustrates how the number of circuits needed to compute $\langle x|x\rangle$ can be reduced. Let's suppose that the Pauli decomposition of a $4\times 4$ matrix reads :

\begin{equation}
A = IX + IZ + XZ + XX + ZX
\end{equation}

\noindent The different terms might have weights which does not affect the process. Following eq. (\ref{eq:hadamard_norm}), these five terms give rise to 10 distinct pairs of Pauli string. Note that the pairs with two identical Pauli strings do not need to be computed as $\langle 0 | V^\dagger A_i^\dagger A_i V|0\rangle=1$. The product of two Pauli operators $P_a^\dagger P_b$ with $P_{a,b} \in \{I,X,Y,Z\}$ respect the following rules:

\begin{table}[h!]
\begin{centering}
\begin{tabular}{ c||c|c|c|c }
 & I & $X$ & $Y$ & $Z$ \\
 \hline \hline
 I$^\dagger$        & I           & $X$  & $Y$   & $Z$ \\
 $X^\dagger$ & $ X$ & I           & $iZ$  &-$iY$ \\
 $Y^\dagger$ & $-Y$ & $iZ$ & -I            &-$iX$\\
 $Z^\dagger$ & $ Z$ & $iY$ & $-iX$ & I \\
\end{tabular}
\caption{Contraction map of single Pauli gates}
\label{table:pauli_contraction}
\end{centering}
\end{table}

Therefore, as seen in Fig. \ref{fig:pauli_simplification}, the Pauli pairs: $(IX)^\dagger (IZ)$, $(XZ)^\dagger (XX)$ are both proportional to the same Pauli string: $IY$. As seen in Fig.  \ref{fig:pauli_simplification}, similar contractions are possible leading to a total of only 7 contracted Pauli pairs. The number of circuits can be further reduced by identifying the contracted Pauli strings that are qubit-wise commutative (QWC) with each other. Two strings are QWC if each Pauli operator in on string commute with the Pauli operator at the same location in the second string. For example $XI$ and $XX$ are QWC but $XI$ and $YX$ not. We can then group the Contracted Pauli pairs in clique where each element of the clique is QWC with all the other element of the same clique. Measuring a single  element of the clique allows to reconstruct the measurements of all the elements in the clique. 

\subsection{Evaluation of $|\langle b | y\rangle|^2$ via classical shadows}

\begin{figure}[h!]
\begin{center}
\begin{tabular}{c}
\includegraphics[scale=0.5]{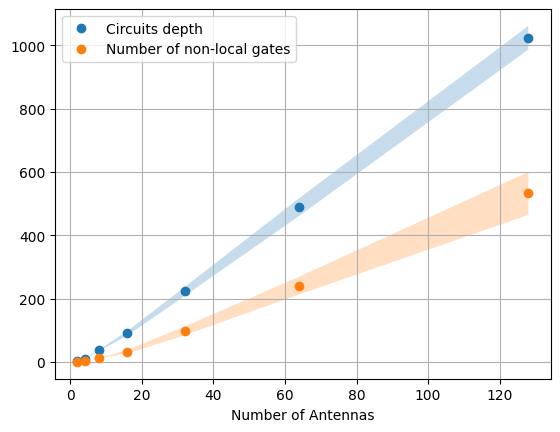} \\
\includegraphics[scale=0.5]{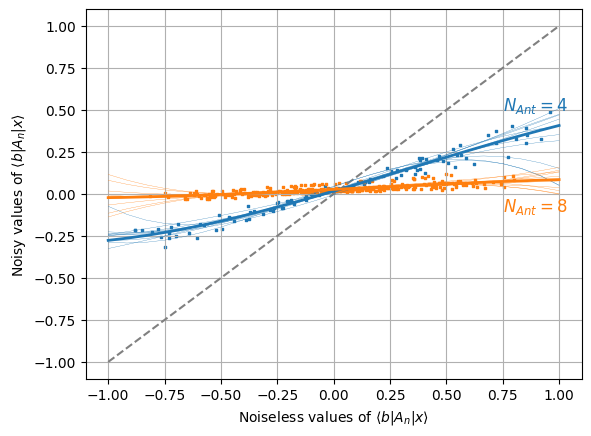}
\end{tabular}
\caption{\label{fig:udepth} Top: Depth and number of non-local gates require to create the $\ket{b}$ vector for different number of antennas in a linear array o the IBMQ-Guadalupe computer.The dots(shaded area) represent mean values(standard deviation) over an ensemble 25 realisations. Bottom: Comparison between noiseless and noisy values (using 100K shots) of the different terms in  $\gamma_n$ (eq. \ref{eq:hadamard_overlap}) for two linear arrays containing 4 and 8 antennas computed via Hadamard Tests. Individual $\gamma_n$ values of 10 different experiments are represented by dots. The thin lines represent third degree polynomial fits of individual experiments while the thick line show the overall fit over all experiments. }
\end{center}
\end{figure}

The Hadamard circuit used to evaluate $|\langle b | y\rangle|^2$  require the creation of a circuit $U$ such as $\ket{b} = U\ket{0}$. As seen in Fig. \ref{fig:udepth},  the synthesis of an arbitrary quantum state is a particularly difficult task that leads quantum circuits whose depth and number of non-local gates exceed the capabilities of current quantum computers even for modest size systems. 

To circumvent this issue we propose to rewrite the terms appearing in eq, (\ref{eq:hadamard_overlap}) as:

\begin{equation}
\gamma_n = \bra{0}V A_n U^\dagger\ket{0} = \sum_i^{2^{N_q}-1}\underbrace{\bra{0}V\ket{i}}_{\textnormal{QST}} \underbrace{\bra{i} A_n \ket{b}}_{\textnormal{classical}}  
\label{eq:qst_overlap}
\end{equation}

The last term of this equation, $\bra{i} A_n \ket{b}$, can be pre-computed classically. The vector $\ket{b}$, encodes the rhs of the linear system and is therefore one of the input of the VQLS algorithm. Since each Pauli matrix $A_n$ is very sparse, with only one non-null element per row, one can compute all the product $A_n\times b$ efficiently and extract the $i$-th component \citep{fastpauli}. 

The remaining term, $\bra{0}V\ket{i}$ can only be evaluated through quantum state tomography (QST). QST allows to reconstruct the full density matrix of the system but usually requires the evaluation of $4^{N_q}$ circuits, which makes it a very computationally intensive task \citep{qst_qiskit}. New approaches, such as Neural Network QST leverage machine learning techniques to limit the computational load require to characterize the quantum state \citep{NeuralQST2018}. Similarly the recently proposed methods based on the evaluation of the classical shadows of the quantum state decreased even further the computational cost of QST \citep{classical_shadows}.  Classical shadows rely on random measurements of the circuits to reconstruct the full density matrix of the system or to evaluate expected values of observables. Several improvements upon the original approach allows to further increase the predictive power of the shadow reconstruction \citep{adapt_shadow, locallybiased_shadow, shadow_noise}. 

\begin{figure}[h!]
\begin{center}
\includegraphics[scale=0.5]{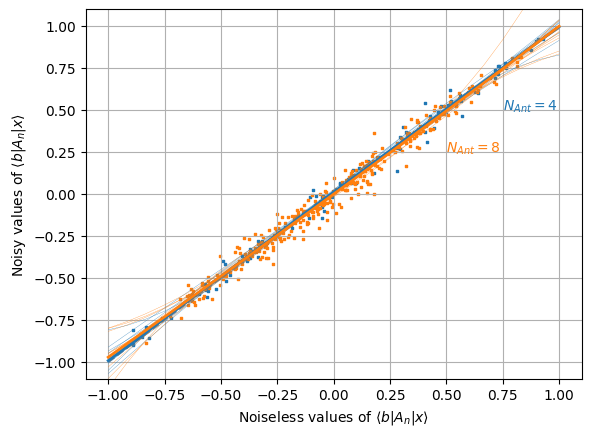}
\caption{\label{fig:hdmr_ovl_qst} Comparison between noiseless and noisy values (using 100K shots) of the different terms in  $\gamma_n$ (eq. \ref{eq:hadamard_overlap}) for two linear arrays containing 4 and 8 antennas computed via classical shadows QST. Individual $\gamma_n$ values of 10 different experiments are represented by dots. The thin lines represent third degree polynomial fit of individual experiments while the thick line show the overall fit over all experiments.}
\end{center}
\end{figure}

The evaluation of $\langle b|y\rangle$ through shadow QST leads to much more robust results in presence of noise. As seen in Fig. \ref{fig:hdmr_ovl_qst}, the noisy values almost exactly match the reference values for 4 and 8 antennas.

\subsection{Evaluation of $\langle y | y\rangle$ via classical shadows}

\begin{figure}[h!]
\begin{center}
\begin{tabular}{c}
\includegraphics[scale=0.5]{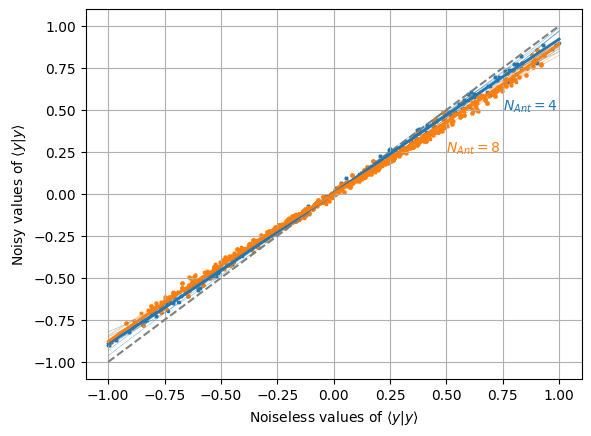} \\
\includegraphics[scale=0.5]{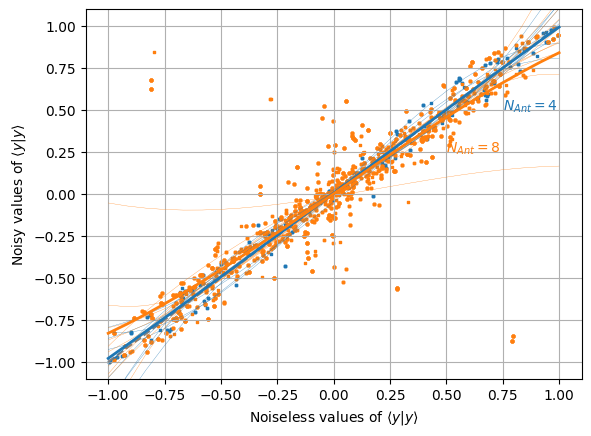}
\end{tabular}
\caption{\label{fig:hdmr_norm_qst} Comparison between noiseless and noisy values (using 100K shots) of the different terms in  $\beta_n$ (eq. \ref{eq:hadamard_norm}) for two linear arrays containing 4 and 8 antennas computed via direct measurements 
(top) and classical shadows QST (bottom). Individual $\beta_n$ values of 10 different experiments are represented by dots. The thin lines represent third degree polynomial fit of individual experiments while the thick line show the overall fit over all experiments.}
\end{center}
\end{figure}

The different terms $\beta_k = \langle x| A_k | x\rangle$ could also be computed via the QST of the variational ansatz: $\beta_k = \textnormal{Tr}(A_k\rho_x)$. However the QWC Pauli strings $A_k$ tends to be highly non-local. The shadow QST approach requires then a very large number of shadows to correctly reconstruct the density matrix. This leads to a significant amount of noise in the values $\beta_k$ with possible outliers (see Fig. \ref{fig:hdmr_norm_qst}). We therefore prefer to use direct measurements of the $A_k$ operators as presented above. The values of these measurements are rather robust in presence of noise (see Fig. \ref{fig:hdmr_norm_qst}). These measurements can also make use of noise mitigation techniques such as Zero Noise Extrapolation to improve the accuracy of the cost function calculation at the expense of increasing the number of quantum circuits to evaluate. 

\section{Solution of the linear systems}
\begin{figure*}
    \centering
    \begin{tabular}{c}
    \includegraphics[scale=0.6]{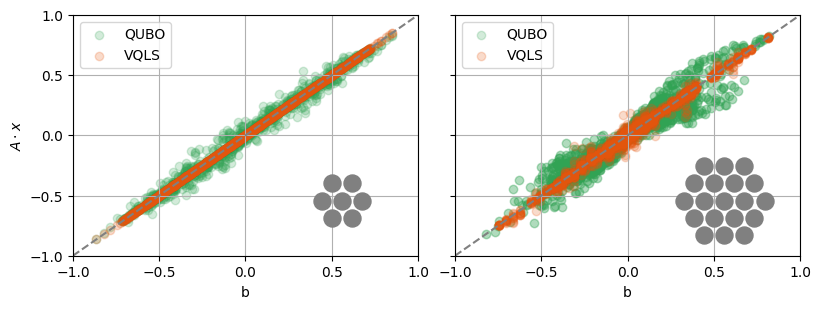} \\  Ideal \\ \hline \\ Realistic \\ 
    \includegraphics[scale=0.6]{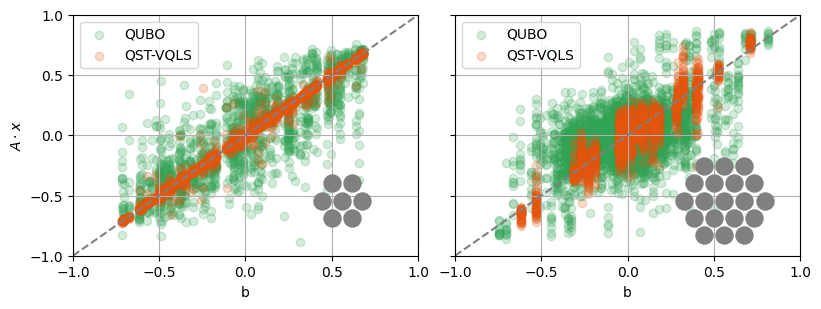} \\
    \includegraphics[scale=0.6]{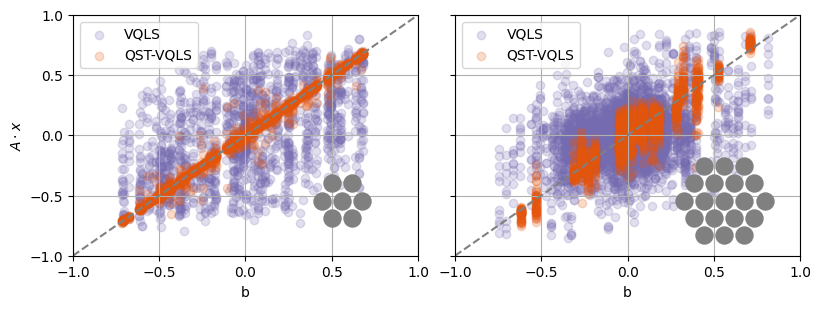}
    \end{tabular}
    \caption{Scatter plots of the values of the product $A \cdot x$ where $A$ is the matrix of the linear system and $x$ the solution provided by the solver and $b$, i.e. the right hand side of the linear system. The results are shown for the ideal case with no noise (top panel) and also accounting for the noise (bottom two panels).}
    \label{fig:residues}
\end{figure*}

Fig. \ref{fig:residues} shows the values of $A\cdot x$ versus the values of $b$ where $A$ is the matrix of the linear system, $x$ the solution provided by the solver and $b$ the right hand side of the linear system. The vectors have here been normalized for better visualization. The results are shown for the ideal case with no noise (top panel) and also accounting for realistic settings of real quantum hardware (bottom two panels). For a perfect solution all the dots will fall onto the diagonal dashed lines. Any deviation from the line therefore indicates an imperfect solution. In the ideal case, the solution provided by the quantum solvers are close to ideal, with minor deviations for the 7 antennas array that increase for the 19 antennas array. When accounting for the noise of the gate-based quantum computers, the QST-VQLS approach provides a satisfying approximation of the linear system solutions. However the bare VQLS approach leads to unsatisfying results with large deviation between the approximate and true solution. The QUBO approach also suffers from comparison with the QST-VQLS approach when it comes to accuracy.  

\end{document}